# Magnetic structures and magnetoelastic coupling of Fe-doped hexagonal manganites LuMn$_{1-x}$Fe$_x$O$_3$ (0 ≤ $x$ ≤ 0.3)

Zhendong Fu,[1] Yinguo Xiao,[2] Harikrishnan S. Nair,[3] Anatoliy Senyshyn,[4,5] Vladimir Y. Pomjakushin,[6] Erxi Feng,[1] Yixi Su,[1] W. T. Jin,[1] and Thomas Brückel[2]

[1] Jülich Centre for Neutron Science JCNS at Heinz Maier-Leibnitz Zentrum MLZ, Forschungszentrum Jülich GmbH, Lichtenbergstraße 1, D-85748 Garching, Germany

[2] Jülich Centre for Neutron Science JCNS and Peter Grünberg Institut PGI, JARA-FIT, Forschungszentrum Jülich GmbH, D-52425 Jülich, Germany

[3] Highly Correlated Matter Research Group, Physics Department, University of Johannesburg P. O. Box 524, Auckland Park 2006, South Africa

[4] Institute for Material Science, Darmstadt University of Technology, D-64287 Darmstadt, Germany

[5] Forschungsneutronenquelle Heinz-Maier Leibnitz FRM-II, Technische Universität München, Licthenbergstraße 1, D-85747 Garching b. München, Germany

[6] Laboratory for Neutron Scattering and Imaging, Paul Scherrer Institut, 5232 Villigen PSI, Switzerland

**Abstract**

We have studied the crystal and magnetic structures of Fe-doped hexagonal manganites LuMn$_{1-x}$Fe$_x$O$_3$ ($x$ = 0, 0.1, 0.2, and 0.3) by using bulk magnetization and neutron powder diffraction methods. The samples crystallize consistently in a hexagonal structure and maintain the space group $P6_3cm$ from 2 to 300 K. The Néel temperature $T_N$ increases continuously with increasing Fe-doping. In contrast to a single $\Gamma_4$ representation in LuMnO$_3$, the magnetic ground state of the Fe-doped samples can only be described with a spin configuration described by a mixture of $\Gamma_3$ ($P6_3'cm'$) and $\Gamma_4$ ($P6_3'c'm$) representations, whose contributions have been quantitatively estimated. The drastic effect of Fe-doping is highlighted by composition-dependent spin reorientations. A phase diagram of the entire composition series is proposed based on the present results and those reported in literature. Our result demonstrates the importance of tailoring compositions in increasing magnetic transition temperatures of multiferroic systems.

## I. INTRODUCTION

Multiferroics are materials in which both ferroelectric and magnetic transitions can occur and where both ordering phenomena coexist in a single phase [1]. The possible coupling between magnetism and ferroelectricity in multiferroics has drawn extensive attention because of the potential technological significance in controlling one order parameter through the other. The hexagonal (*h*-) manganites $R$MnO$_3$ ($R$ = rare earth) are an interesting group of multiferroics which exhibit a rich variety of physical phenomena [2,3]. Although the ferroelectric transition temperature ($T_F$ ~1000 K) of *h*-$R$MnO$_3$ is much higher than its antiferromagnetic (AFM) transition temperature ($T_N$ ~100 K) [3], evidence for coupling between magnetic and electric dipole moments has been revealed by means of dielectric constant measurements and high-resolution neutron diffraction [4,5].

In the ferroelectric state *h*-$R$MnO$_3$ are crystallized in the hexagonal structure with the space group $P6_3cm$ [6]. As shown in Fig. 1(a), each Mn atom and its five adjacent oxygen atoms form a MnO$_5$ bipyramid, where two oxygen atoms are at the apexes and three oxygen atoms are in the equatorial plane of the MnO$_5$ bipyramid. The corner-sharing MnO$_5$ bipyramids form a triangular lattice in the *a-b* plane of the hexagonal structure and are well separated from each other along the *c* axis by the plane of $R$ ions. Pronounced magnetic frustration has been observed in *h*-$R$MnO$_3$, arising from the 120º triangular lattice of antiferromagnetically coupled Mn$^{3+}$ ions. The Mn-O-O-Mn superexchange interaction between adjacent triangular planes ($z$ = 0 and $z$ = 1/2) is found much weaker than the in-plane Mn-O-Mn superexchange interaction [3,7].

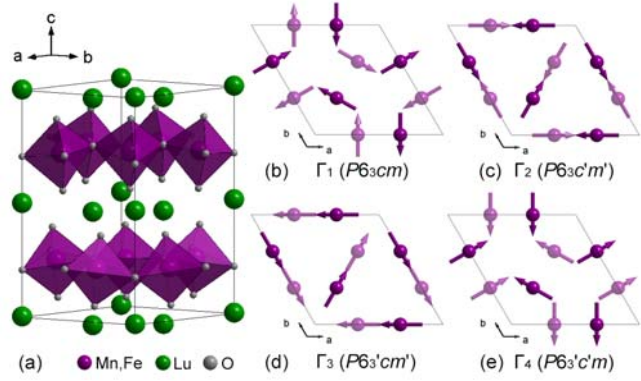

FIG. 1. (Color online) (a) Schematic of the crystal structure of $LuMn_{1-x}Fe_xO_3$; (b)–(e) four possible 1D magnetic representations of $LuMn_{1-x}Fe_xO_3$: $\Gamma_1$ ($P6_3cm$), $\Gamma_2$ ($P6_3c'm'$), $\Gamma_3$ ($P6_3'cm'$), and $\Gamma_4$ ($P6_3'c'm$). The lighter and darker symbols denote Mn/Fe atoms displaced along $c$ by 1/2 of the unit cell.

The $Mn^{3+}$ moments in $h$-$R$MnO$_3$ order below $T_N$ in a non-collinear spin structure with a 120º angular difference between neighboring spins in the $a$-$b$ plane [4,5]. However, precise determination of the magnetic structure of $h$-$R$MnO$_3$ is often nontrivial due to the existence of homometric spin configurations and limited instrumental resolution. For example, as one of the most intensively studied $h$-$R$MnO$_3$, YMnO$_3$ has an AFM structure with $\mathbf{k} = 0$ vector below $T_N$ as suggested by neutron diffraction measurements. According to the theoretical analysis using magnetic group theory, altogether six magnetic structures are found to be possible: four 1D ($\Gamma_1$, $\Gamma_2$, $\Gamma_3$, and $\Gamma_4$) and two 2D ($\Gamma_5$ and $\Gamma_6$) irreducible representations [8]. The four 1D models are shown in Fig. 1(b)–(e). $\Gamma_1$ ($P6_3cm$) and $\Gamma_2$ ($P6_3c'm'$) have antiparallel coupling between $z = 0$ and $z = 1/2$ moments, while $\Gamma_3$ ($P6_3'cm'$) and $\Gamma_4$ ($P6_3'c'm$) have parallel coupling. The moment on the [100] axis is perpendicular to the [100] axis in $\Gamma_1$ and $\Gamma_4$, and parallel in $\Gamma_2$ and $\Gamma_3$. Earlier neutron powder diffraction (NPD) studies have proposed that the magnetic structure of YMnO$_3$ is either $\Gamma_1$ or $\Gamma_3$, which were indistinguishable within the experimental resolution [9,10]. The second harmonic generation results agree better with the $\Gamma_3$ magnetic symmetry [11,12], while $\Gamma_1$ is also favored in other literature [13,14]. The later polarimetric neutron study has suggested a $\Gamma_6$ ($P6_3'$) magnetic symmetry with Mn moments inclined at 11º with respect to $\Gamma_3$ [15]. The key parameter in selecting a given magnetic structure or the occurrence of a spin reorientation transition in hexagonal manganites has been attributed to the position $x$ of Mn ions within the triangular plane with reference to a critical threshold of 1/3 using high-resolution NPD and inelastic neutron scattering methods [16,17].

Despite the similarities in physical properties with YMnO$_3$, LuMnO$_3$ in AFM phase has been described by either a $\Gamma_4$ representation [11,18,19], or a $\Gamma_2$ representation [20]. Doping Lu at Y site of YMnO$_3$ introduces a continuous variation in the magnetic structure from $\Gamma_3$ to $\Gamma_4$ representation, which can be explained in terms of the chemical pressure effects due to doping [18]. It is also interesting to explore the effect of the magnetic doping at Mn site on the magnetic structure. A single phase of $h$-YMn$_{1-x}$Fe$_x$O$_3$ has been obtained for $x \leq 0.3$ [21]. 10% Fe doping at the Mn site of YMnO$_3$ results in a decrease of $T_N$ from 75 K to 60 K and a spin reorientation transition from $\Gamma_3$ to $\Gamma_3$ + $\Gamma_4$ at $T_{SR} \sim 35$ K [22]. In a recent NPD study of Fe-doped (up to 10%) $h$-YMn$_{1-x}$Fe$_x$O$_3$, the magnetic ground state has been found to change from a highly frustrated ($\Gamma_1$ in YMnO$_3$) to a lowly frustrated ($\Gamma_2$ in YMn$_{0.9}$Fe$_{0.1}$O$_3$) magnetic structure, via a mixed [($\Gamma_1$ + $\Gamma_2$) in YMn$_{0.95}$Fe$_{0.05}$O$_3$] configuration [14]. The Fe doping in LuMnO$_3$ has also attracted attention due to the improved magnetic properties of $h$-LuFeO$_3$ as compared with those of $h$-LuMnO$_3$. A single phase of solid solution has been achieved in the half doped LuMn$_{0.5}$Fe$_{0.5}$O$_3$ [2,23]. Indeed the $T_N$ ($\sim 110$ K) of LuMn$_{0.5}$Fe$_{0.5}$O$_3$ is higher than that of LuMnO$_3$ by about 20 K. The magnetic ground state of LuMn$_{0.5}$Fe$_{0.5}$O$_3$ has been described by a single magnetic representation $\Gamma_1$ or $\Gamma_3$ [2,23]. The crystal and magnetic structures of the Fe-rich compounds, LuMn$_{1-x}$Fe$_x$O$_3$ ($x \geq 0.5$), have been investigated using powder diffraction and inelastic neutron scattering in Ref. 23, which will be revisited later in the discussion. In the present work, we report the effect of Fe-doping (up to 30%) on the magnetic structure and magnetic properties of $h$-LuMnO$_3$.

## II. EXPERIMENTAL DETAILS

The polycrystalline samples used in this paper were synthesized by the standard solid-state reaction method [2,24]. Subsequent x-ray diffraction (XRD) was done on a Huber x-ray diffractometer (Huber G670) with Cu-$K\alpha$ radiation at room temperature. The XRD patterns (not shown) for all as-prepared samples indicate no trace of impurity phases and were indexed in a hexagonal structure with space group $P6_3cm$. Magnetization data were taken using a Quantum Design Dynacool physical property measurement system. The temperature-dependent NPD experiments were carried out on the high-resolution diffractometer for thermal neutrons HRPT [25] at the SINQ spallation source of Paul Scherrer Institute PSI (Villigen, Switzerland), and on the high-resolution powder diffractometer SPODI [26] at the Heinz Maier-Leibnitz Zentrum (MLZ), Garching, Germany. During measurement, HRPT was running in high-intensity mode with the wavelength $\lambda = 1.886$ Å. The wavelength used on SPODI was 1.5483 Å. The NPD data are normalized by the spectrum of a vanadium standard. The irreducible representations for magnetic structure were obtained using SARAh package [27]. The nuclear and magnetic structure refinements were performed using Rietveld method [28] with FULLPROF suite [29].

## III. RESULTS AND ANALYSIS

As shown in Fig. 2(a), the temperature dependences of molar magnetization $M$ of LuMn$_{1-x}$Fe$_x$O$_3$ ($x$ = 0, 0.1, 0.2, and 0.3) were measured under an external field of 1 T from 5 to 310 K. All four samples show a magnetic phase transition in the range between 90 and 115 K, corresponding to the Néel transition $T_N$ from the paramagnetic state to the long-range AFM ordered state. Considering also the recently published work on LuMn$_{0.5}$Fe$_{0.5}$O$_3$ [2], we find that the $T_N$ of LuMn$_{1-x}$Fe$_x$O$_3$ increases from 92 to 112 K when the Fe-doping ratio increases from 0 to 0.5. Similar to LuMn$_{0.5}$Fe$_{0.5}$O$_3$ [2], LuMn$_{0.7}$Fe$_{0.3}$O$_3$ exhibits another phase transition below $T_N$ at $T_{SR}$ ~55 K. We tentatively attribute it to a spin reorientation transition, which will be discussed later in this paper. The inverse susceptibility $H/M$ *vs.* temperature $T$ is plotted in Fig. 2(b). The fit with the Curie-Weiss law to the data above 250 K reveals a negative Curie-Weiss temperature $\theta_{CW}$ in all samples, indicative of predominant AFM coupling among magnetic moments. The effective paramagnetic moment $\mu_{eff}$ of each sample is calculated from the Curie constant and listed in Table I. The values of the frustration parameter, $f = |\theta_{CW}|/T_N$, for the samples are around or larger than 5, suggesting that magnetic frustration may exist in the samples. The triangular-lattice arrangement of the Mn atoms has been considered as the main source for the geometrical magnetic frustration in $h$-$R$MnO$_3$ [2,13,15,18,30]. However, we should point out that the difference between $\theta_{CW}$ and $T_N$ can also be due to the low-dimensionality of the magnetic lattice because the intra-plane coupling is much stronger than the inter-plane coupling. The deviation from Curie-Weiss behavior in the samples takes place at temperatures much higher than $T_N$ due to the onset of magnetic 2D short-range correlations, as evidenced by the magnetic diffuse scattering which will be discussed later. The obtained $T_N$, $T_{SR}$, $\theta_{CW}$, and $\mu_{eff}$ are summarized in Table I. The field dependence of magnetization (not shown) has been taken from all samples at 5 K. The absence of hysteresis rules out a ferromagnetic contribution to the magnetic susceptibility.

TABLE I. The Curie-Weiss temperature $\theta_{CW}$, the Néel temperature $T_N$, the frustration parameter $f$, the temperature of spin reorientation $T_{SR}$, and the effective paramagnetic moment $\mu_{eff}$ of LuMn$_{1-x}$Fe$_x$O$_3$ ($x \leq 0.5$). The values for LuMn$_{0.5}$Fe$_{0.5}$O$_3$ are taken from Ref. 2.

|  | LuMnO$_3$ | LuMn$_{0.9}$Fe$_{0.1}$O$_3$ | LuMn$_{0.8}$Fe$_{0.2}$O$_3$ | LuMn$_{0.7}$Fe$_{0.3}$O$_3$ | LuMn$_{0.5}$Fe$_{0.5}$O$_3$ |
|---|---|---|---|---|---|
| $|\theta_{CW}|$ (K) | 776(3) | 542(7) | 510(4) | 552(6) | 946 |
| $T_N$ (K) | 92(1) | 102(1) | 105(1) | 109(1) | 112 |
| $f$ | 8.4(2) | 5.3(1) | 4.8(1) | 5.1(1) | 8.5 |
| $T_{SR}$ (K) | --- | --- | --- | 55(1) | 55 |
| $\mu_{eff}$ ($\mu_B$) | 4.88(5) | 5.17(6) | 5.15(5) | 5.27(4) | 5.41(4) |

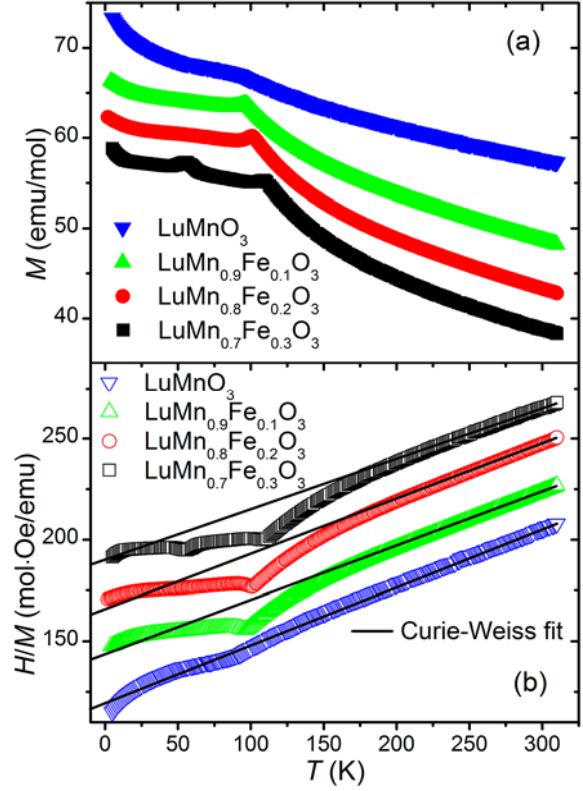

FIG. 2. (Color online) (a) Molar magnetization for $LuMn_{1-x}Fe_xO_3$ ($x$ = 0, 0.1, 0.2, and 0.3) as a function of temperature between 5 and 310 K; (b) temperature dependence of the inverse susceptibility with the fit using Curie-Weiss law. The curves are shifted vertically for clarity.

In order to investigate the Fe-doping effect on the crystal and magnetic structures of $LuMnO_3$, the NPD patterns for $LuMn_{1-x}Fe_xO_3$ ($x$ = 0, 0.1, and 0.3) were recorded from 2 to 300 K on HRPT, and those for $x$ = 0.2 were recorded from 5 to 300 K on SPODI. Representative NPD patterns measured at 5 and 300 K on each sample are shown in Fig. 3 as a function of $Q$ ($\equiv 4\pi\sin(\theta)/\lambda$), along with Rietveld refinement results. It is further confirmed from the well-refined neutron diffraction patterns that our samples are of single phase with little trace of impurities. In the temperature range of interest from 2 to 300 K, no symmetry change has been observed and the nuclear structure of all samples belongs to the space group $P6_3cm$.

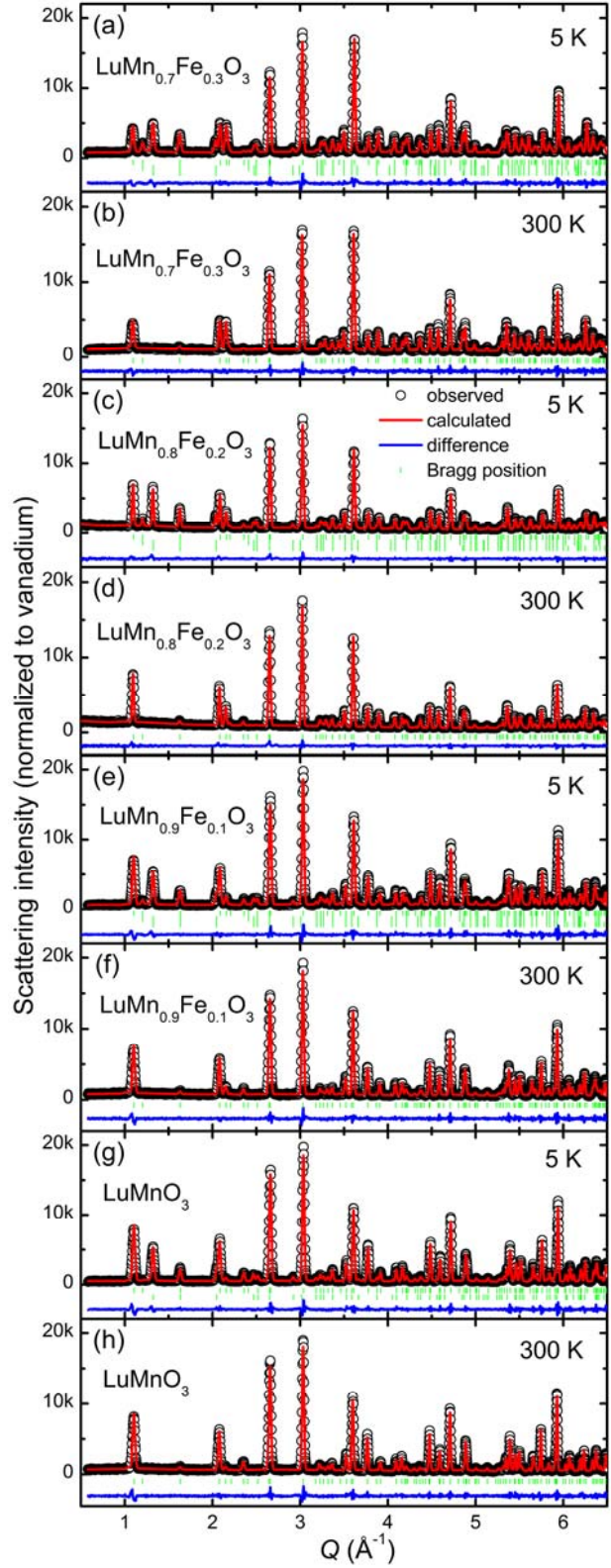

FIG. 3. (Color online) Neutron powder diffraction patterns for LuMn$_{1-x}$Fe$_x$O$_3$ ($x$ = 0, 0.1, 0.2, and 0.3) measured at 5 K and 300 K, along with Rietveld refinement results.

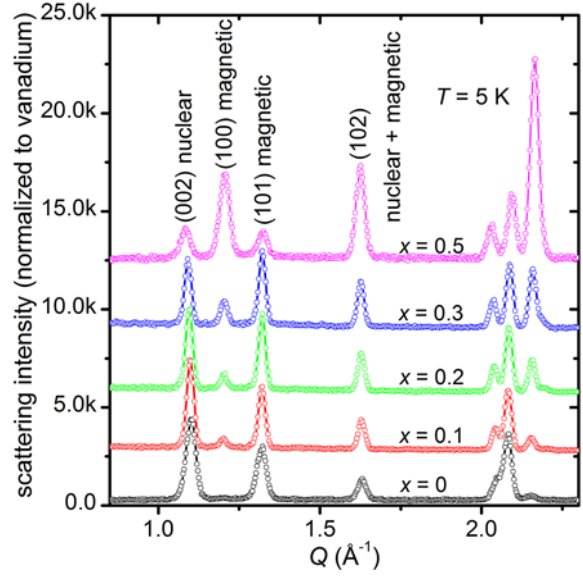

FIG. 4. (Color online) Neutron powder diffraction patterns over a $Q$ range of 0.8 – 2.3 Å$^{-1}$ for LuMn$_{1-x}$Fe$_x$O$_3$ ($x$ = 0, 0.1, 0.2, 0.3, and 0.5) measured at 5 K. The patterns are shifted vertically for clarity. The data for LuMn$_{0.5}$Fe$_{0.5}$O$_3$ are taken from Ref. 2.

All four samples show clear magnetic reflections below $T_N$. Fig. 4 shows the low-angle part of the NPD patterns recorded at 5 K. For comparison, the data of LuMn$_{0.5}$Fe$_{0.5}$O$_3$ from our previous publication is also shown [2]. It is obvious that Fe doping at the Mn-site has dramatic influence on the magnetic structure of $h$-LuMnO$_3$. The magnetic propagation vector is found to be $\mathbf{k}$ = (0, 0, 0) for all samples. In order to fit the data, four 1D irreducible magnetic representations, $\Gamma_1$ ($P6_3cm$), $\Gamma_2$ ($P6_3c'm'$), $\Gamma_3$ ($P6_3'cm'$), and $\Gamma_4$ ($P6_3'c'm$), were generated by SARAh package. The calculated (100) nuclear reflection is so weak that the observed (100) reflection is considered to be of purely magnetic origin. Its intensity increases with increasing Fe-doping. Because the magnetic (100) reflection is only allowed in the magnetic structure $\Gamma_1$ or $\Gamma_3$, it is suggested that the magnetic phases of the Fe-doped samples must contain contributions from $\Gamma_1$ or $\Gamma_3$ representation, whose percentage seems to increase with increasing Fe-doping. Satisfying fit to the data for LuMnO$_3$ can be obtained by using either $\Gamma_2$ or $\Gamma_4$. The data for LuMn$_{1-x}$Fe$_x$O$_3$ ($x$ = 0.1, 0.2, and 0.3), on the other hand, cannot be fitted properly by any single 1D irreducible representation. The best fit requires a combination of irreducible representations, ($\Gamma_1+\Gamma_2$) or ($\Gamma_3+\Gamma_4$). The possible combinations, ($\Gamma_1+\Gamma_4$) or ($\Gamma_2+\Gamma_3$), are excluded because they result in unrealistic magnetic structures with different ordered moments at $z$ = 0 and 1/2 planes [31]. We adopt the combination ($\Gamma_3+\Gamma_4$) to fit the NPD data for the Fe-doped samples in this paper, but the fit with ($\Gamma_1+\Gamma_2$) can give equally good results. We have achieved good agreement factors in the refinement of the data for all samples: $\chi^2$ < 3.5%, $R_p$ < 5.5%, and $R_{wp}$ < 7.5%.

Fig. 5 shows the temperature dependence of the refined lattice constants, $a$ and $c$, and the unit-cell volume $V$ for LuMn$_{1-x}$Fe$_x$O$_3$ ($x$ = 0, 0.1, 0.2, and 0.3). The $a$ parameter decreases as temperature decreases from 300 to 30 K and remains nearly constant below 30 K. The thermal evolution of the unit-cell volume $V$ is similar to that of the $a$ parameter. The expansions of $a$ and $V$ from 5 to 300 K are about 0.015 Å and 1.7 Å$^3$ in all samples. The $c$ parameter basically shows a thermal contraction when approaching $T_N$ from below, but the change of the value of $c$ is less than 0.003 Å within the investigated temperature range. It is clear that the thermal evolution in $a$-$b$ plane is more significant than in $c$ direction. As the Fe-doping ratio increases from 0 to 0.3, the $a$ parameter at base temperature decreases by about 0.02 Å, while the $c$ parameter and the unit-cell volume increase by about 0.11 Å and 1.0 Å$^3$, respectively.

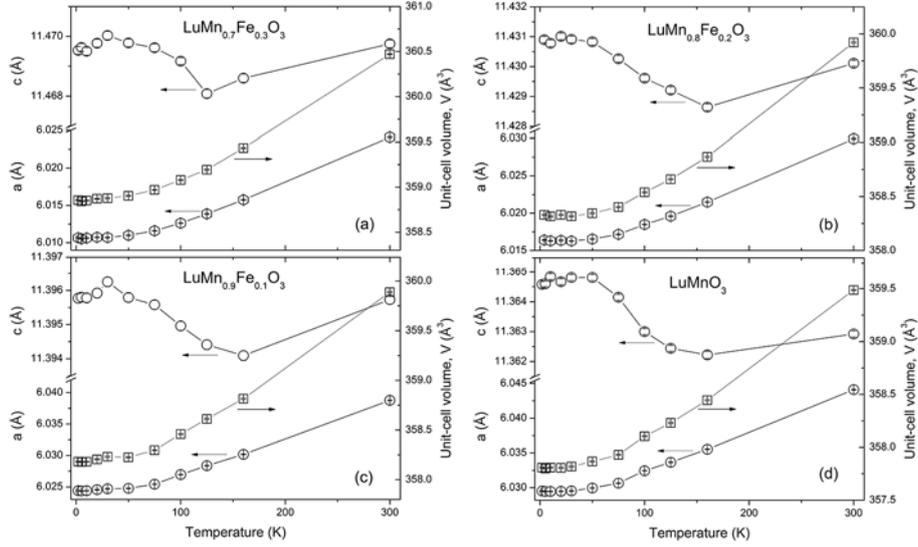

FIG. 5. Temperature dependence of the lattice constants, $a$ and $c$, and the unit-cell volume for LuMn$_{1-x}$Fe$_x$O$_3$ ($x$ = 0, 0.1, 0.2, and 0.3).

The temperature dependence of the ordered magnetic moment, $m_{ord}(T)$, is plotted in Fig. 6(a). The ordered moments of the samples saturate between 3.2 and 3.5 $\mu_B$ below 20 K and show no clear dependence on the Fe-doping ratio. Note that the maximum ordered spin-only moments of Mn$^{3+}$ and Fe$^{3+}$ are 4 and 5 $\mu_B$, respectively. The reduction of the ordered moments with respect to the spin-only values indicates that intrinsic quantum fluctuation still exists in the ordered phase. The $m_{ord}(T)$ of LuMn$_{0.8}$Fe$_{0.2}$O$_3$ within the temperature range between 50 K and 100 K is fitted with a power law, $m_{ord}(T) \sim (1-T/T_N)^\beta$. As shown by the dash-and-dot line in Fig. 6(a), the fit yields an ordering temperature of $T_N$ = 100.8 ± 1.2 K and $\beta$ = 0.21 ± 0.03. The value of $\beta$ is smaller than that expected for a 3D Heisenberg system, in agreement with the layered nature of the magnetic lattice in our samples.

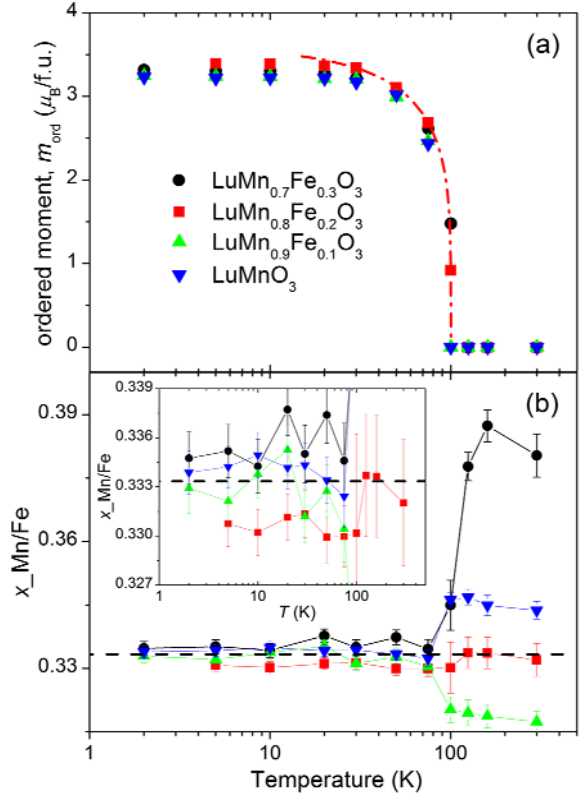

FIG. 6. (Color online) (a) Temperature dependence of the ordered moments for $LuMn_{1-x}Fe_xO_3$ ($x = 0$, 0.1, 0.2, and 0.3). Dash-and-dot line represents a fit to the data for $LuMn_{0.8}Fe_{0.2}O_3$ with a power law. (b) Temperature dependence of the $x$ position of Mn/Fe atoms, $x$_Mn/Fe, for $LuMn_{1-x}Fe_xO_3$ ($x = 0$, 0.1, 0.2, and 0.3). The inset of (b) depicts the zoom-in around $x$_Mn/Fe = 1/3.

The $x$ position of Mn ($x$_Mn) and Mn–O bond distance are often considered as key parameters which describe the doping effect and determine the stability of the magnetic phases in $h$-$R$MnO$_3$ [16,17]. When $x$_Mn = 1/3, the Mn ions form an ideal triangular lattice in the $a$-$b$ plane. The Mn-Mn exchange paths in the $a$-$b$ plane, as well as the inter-plane paths, are equivalent [17]. Deviation of $x$_Mn from 1/3 leads to different intra-plane and inter-plane magnetic exchange interactions, which can be comprehended through the change in Mn–O bond distances. Such displacements of Mn atoms have strong correlation with the magnetic structure and have been observed in various $h$-$R$MnO$_3$, e.g., $x$_Mn = 0.342 for YMnO$_3$ [16] and 0.331 for LuMnO$_3$ [18] at 10 K. The temperature dependence of $x$_Mn/Fe in $LuMn_{1-x}Fe_xO_3$ is shown in Fig. 6(b). For $LuMn_{0.7}Fe_{0.3}O_3$ the temperature dependence of $x$_Mn/Fe is similar with the one for $LuMn_{0.5}Fe_{0.5}O_3$ [2]. Below $T_N$, $x$_Mn/Fe is nearly constant and remains above the 1/3 threshold. Then a sharp increase of $x$_Mn/Fe occurs at $T_N$ by a surprisingly large value of ~0.05, corresponding to a relative shift of almost 15%. As far as we know, this is the largest shift of $x$_Mn ever observed in $h$-$R$MnO$_3$ when the temperature crosses $T_N$. For $LuMn_{0.8}Fe_{0.2}O_3$, $x$_Mn/Fe first remains nearly constant below 1/3 threshold below $T_N$ and then increases slightly to the 1/3 threshold above $T_N$. In LuMn$_{0.9}$Fe$_{0.1}$O$_3$, on the other hand, $x$_Mn/Fe remains around 1/3 below $T_N$ and decreases upon heating above $T_N$. In the case of LuMnO$_3$, $x$_Mn is 0.334 at 2 K. Then $x$_Mn increases with increasing temperature to about 0.345. The thermal evolution of $x$_Mn in our LuMnO$_3$ sample is a little different from that reported in literature [16]. But we also notice that the samples with the same composition may have slight difference in structural properties due to the difference in synthesizing procedures [32,33].

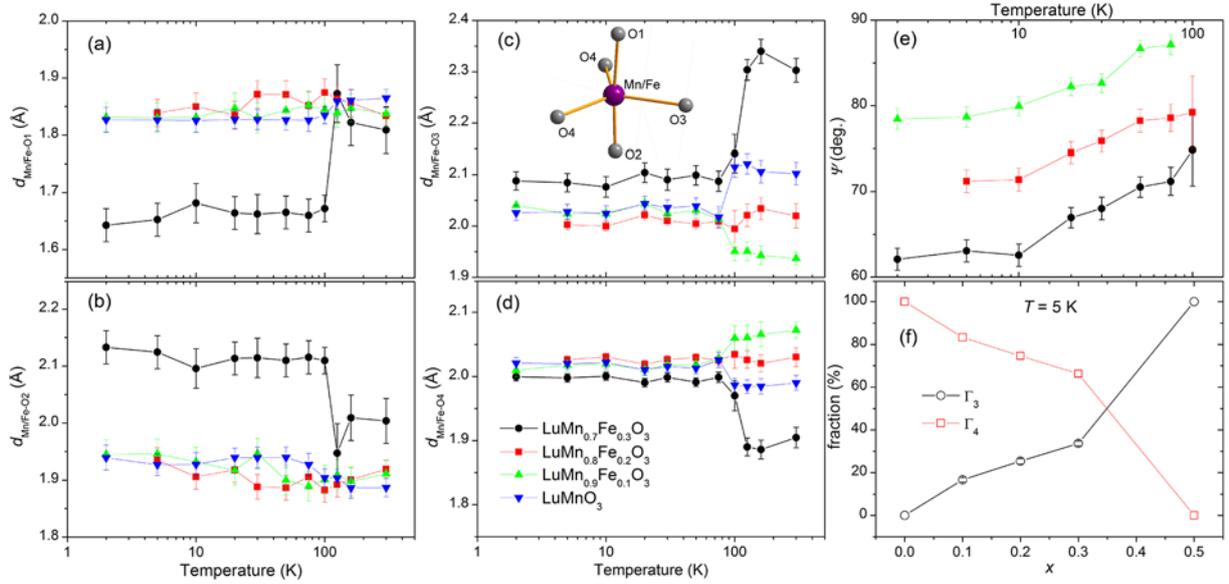

FIG. 7. (Color online) Temperature dependence of the bond lengths, Mn/Fe–O1 (a), Mn/Fe–O2 (b), Mn/Fe–O3 (c), and Mn/Fe–O4 (d) for LuMn$_{1-x}$Fe$_x$O$_3$ ($x$ = 0, 0.1, 0.2, and 0.3), the inset of (c) shows the oxygen coordination environment of Mn/Fe ions; (e) temperature evolution of the angle ($\psi$) between the direction of the Mn/Fe moment at the position ($x$_Mn/Fe, 0, 0) and the $a$ axis in the Fe-doped samples; (f) the evaluated fractions of $\Gamma_3$ and $\Gamma_4$ as a function of the Fe concentration $x$.

The temperature dependence of the Mn/Fe–O bond distances are plotted in Fig. 7(a)–(d). A schematic view of the oxygen coordination environment of Mn/Fe ions is given in the inset of Fig. 7(c). Below $T_N$ LuMn$_{0.7}$Fe$_{0.3}$O$_3$ has a difference as large as 0.45 Å between Mn/Fe–O1 and Mn/Fe–O2. For other samples, the bond lengths of Mn/Fe–O1 and Mn/Fe–O2 are close and remain nearly constant in the investigated temperature range. As seen in Fig. 7(c) and (d), the in-plane Mn/Fe–O3 and Mn/Fe–O4 bond lengths of LuMn$_{0.7}$Fe$_{0.3}$O$_3$ deviate significantly above $T_N$ due to the large shift of Mn/Fe atom. The in-plane bond lengths of LuMn$_{0.9}$Fe$_{0.1}$O$_3$ and LuMnO$_3$ show slight differences above $T_N$, while those of LuMn$_{0.8}$Fe$_{0.2}$O$_3$ are nearly equal in the investigated temperature range.

By mixing 1D irreducible representation $\Gamma_3$ and $\Gamma_4$, we obtain a 2D magnetic model with the space group $P6_3'$,

where there is an angle ($\psi$) between the direction of the Mn/Fe moment at the position ($x\_Mn/Fe$, 0, 0) and the $a$ axis, i.e., $\psi = 0°$ for $\Gamma_3$ and $\psi = 90°$ for $\Gamma_4$. In order to investigate the evolution of $\psi$ with temperature, we plot the temperature dependence of $\psi$ for LuMn$_{1-x}$Fe$_x$O$_3$ ($x$ = 0.1, 0.2, 0.3) in Fig. 7(e). It is interesting to note that for all three samples, $\psi$ increases with increasing temperature, corresponding to an overall rotation of the moments from $\Gamma_3$ to $\Gamma_4$. This is consistent with the recent first-principles calculations of the magnetic energies for LuMnO$_3$, which indicate that $\Gamma_4$ spin configuration is a little higher in energy than $\Gamma_3$ [34]. Fig. 7(e) also suggests that $\psi$ decreases with increasing Fe-doping. This doping effect can be seen clearly in Fig. 7(f), where the percentages of $\Gamma_3$ and $\Gamma_4$ configurations are plotted as a function of Fe concentration. When the Fe-doping ratio increases from 0 to 0.5, the fraction of $\Gamma_4$ decreases from 100% to 0. Therefore the observed change of magnetic structure upon Fe doping is interpreted as a chemically-driven spin reorientation in the *a-b* plane, while the spin reorientation with increasing temperature can be regarded as a thermally-driven reorientation of the moments towards the configuration with a higher energy.

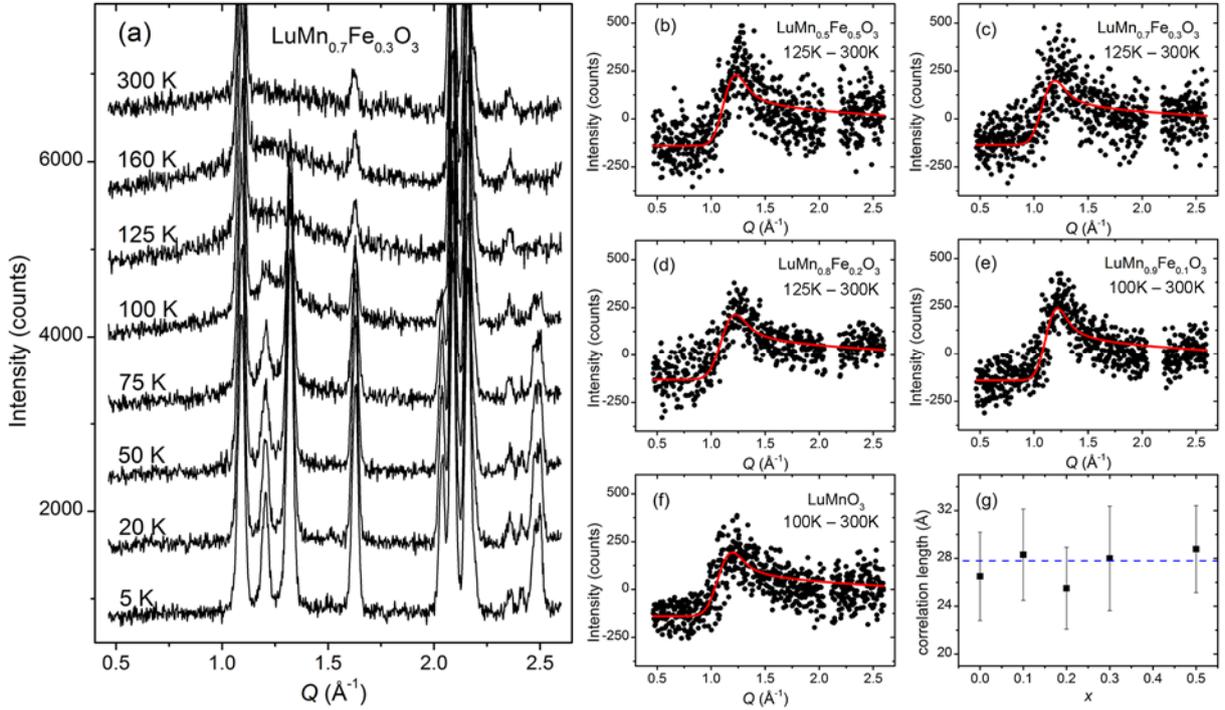

FIG. 8. (Color online) (a) Selected NPD patterns for LuMn$_{0.7}$Fe$_{0.3}$O$_3$ at various temperatures; (b)–(f) magnetic diffuse scattering for LuMn$_{1-x}$Fe$_x$O$_3$ ($0 \leq x \leq 0.5$) obtained by subtracting the NPD data collected at 300 K from those collected at a temperature close to $T_N$. The red lines are the fit with 2D Warren function; (g) the 2D correlation lengths as a function of the concentration of Fe at temperatures close to $T_N$.

Besides the magnetic reflections below $T_N$, strong magnetic diffuse scattering has been revealed in our samples.

The magnetic diffuse scattering has been widely observed in $h$-$R$MnO$_3$ and considered as the evidence of geometrical spin frustration in literature [18,30]. The representative NPD patterns at various temperatures for LuMn$_{0.7}$Fe$_{0.3}$O$_3$ are plotted in Fig. 8(a). The diffuse scattering of other samples shows similar temperature dependence with the one of LuMn$_{0.7}$Fe$_{0.3}$O$_3$. As seen in Fig. 8(a), clear magnetic diffuse scattering has emerged already at 160 K around the position of magnetic (100) reflection (~1.2 Å$^{-1}$) and is most pronounced at temperatures around $T_N$. As the spins start to order below $T_N$, this diffuse peak becomes subdued with further cooling from $T_N$. In order to extract the diffuse scattering, we subtracted off the diffraction pattern taken at 300 K, where the spins should be in a paramagnetic state, from the patterns taken at temperatures below 300 K. The diffuse scattering patterns at temperatures close to $T_N$ for LuMn$_{1-x}$Fe$_x$O$_3$ ($0 \leq x \leq 0.5$) are plotted as a function of $Q$ in Fig. 8(b)–(f). The NPD data for LuMn$_{0.5}$Fe$_{0.5}$O$_3$ is taken from Ref. 2. A common feature about these diffuse peaks is the asymmetric shape (a fast rise at low $Q$ and a slow fall towards high $Q$), which is characteristic of a 2D short-range order and agrees well with the profile of the weakly coupled Mn/Fe layers in LuMn$_{1-x}$Fe$_x$O$_3$. The magnetic diffuse scattering can then be described analytically by a modified Warren function for 2D magnetic correlations as follows [35-37],

$$P(Q)/F_m^2 = KmF_{hk}^2 \frac{1+\left(1-2(\lambda Q)^2/(4\pi)^2\right)^2}{2(\lambda Q/4\pi)^{3/2}} \left(\frac{\xi}{\lambda\sqrt{\pi}}\right)^{1/2} F(a) + C, \quad (1)$$

with

$$a = \xi(Q-Q_0)/(2\sqrt{\pi}), \quad (2)$$

and

$$F(a) = \int_0^{10} \exp[-(x^2-a)^2]dx. \quad (3)$$

$F_m$ is the magnetic form factor of magnetic ions. $K$ is a scaling factor. $m$ is the multiplicity of the 2D reflection ($hk$) with the magnetic structure factor $F_{hk}$. $\lambda$ is the wavelength of neutrons. $\xi$ is the 2D spin-spin correlation length. $C$ is a constant accounting for the subtraction of magnetic form factor. $Q_0$ is the position of the reflection ($hk$). The fitting results are plotted in Fig. 8(b)–(f). The 2D correlation length $\xi$, estimated from the above fitting, are summarized as a function of the Fe concentration in Fig. 8(g). The $\xi$ for all samples are around 28 Å. We do not see a regular pattern of the correlation length upon Fe-doping, which could be hindered by the large error bars due to the noisy diffuse signal obtained by subtraction. Polarized neutron scattering technique will be necessary to characterize the correlation lengths more precisely [38-40]. The fit with the Warren function shows that the diffuse scattering is most likely due to the short-range order which originates from the strong exchange coupling in $a$-$b$ planes. At $T_N$ the 3D magnetic order occurs when the *a-b* planes are locked in with respect to each other due to the inter-plane exchange interactions [41]. Our analysis suggests that the low dimensionality of the magnet lattice plays a more important role than the geometrical frustration in shaping the magnetic behaviors of LuMn$_{1-x}$Fe$_x$O$_3$.

## IV. DISCUSSION

Our structural characterization confirms that the $P6_3cm$ hexagonal symmetry, common in *h-R*MnO$_3$ family, is preserved in LuMn$_{1-x}$Fe$_x$O$_3$ ($0 \leq x \leq 0.3$), within the temperature range of 2–300 K. Strong magneto-elastic coupling occurs through the large atomic displacements, which is rather pronounced in the samples with high Fe-doping ratio. With increasing Fe-doping, we observed an increase of the lattice constant *c* and the unit-cell volume, but a decrease of the lattice constant *a*. The change of *c* is stronger than the change of *a* upon Fe-doping, suggesting that the equivalent chemical pressure introduced by Fe-doping is stronger in *c* axis than in the basal plane axis. On the contrary, Y-doping to the Lu site causes stronger chemical pressure effect in the basal plane [18]. The expansion of *c* with Fe-doping mainly reflects the increased buckling of Mn/Fe-O$_5$ polyhedra [14]. Applying either chemical or physical pressure may produce a subtle change in the magnetic easy axis and in turn a spin reorientation in YMnO$_3$ [18,42]. Although no change in the symmetry of the triangular AFM state of LuMnO$_3$ was observed at a high pressure of 6 GPa [20], spin reorientation in *a-b* plane may occur by either Y [18] or Fe (this work) doping in LuMnO$_3$. We have found that the rotation angle $\psi$ decreases from 90° to 0° as the Fe-doping ratio increases from 0 to 0.5. We also notice in a recent work [23] on Mn-doped *h*-LuFeO$_3$, the ground-state spin configuration shows a rotation from $\Gamma_2$ to $\Gamma_1$ (or $\Gamma_3$) via an intermediate representation ($\Gamma_2 + \Gamma_1/\Gamma_3$) as the Mn concentration increases from 0.25 to 0.5. These results highlight the composition-driven spin reorientation owing to the chemical disturbance to the single phase compound.

We observed a small peak at 55 K in the temperature dependent magnetization of LuMn$_{0.7}$Fe$_{0.3}$O$_3$ in Fig. 2(a). Such a peak also exists in LuMn$_{0.5}$Fe$_{0.5}$O$_3$ [2,23]. We attribute this peak to a spin-reorientation behavior, similar to the ones found in LuFe$_{1-x}$Mn$_x$O$_3$ ($x$ = 0.25 and 0.33) [23], LuFeO$_3$ [43], HoMnO$_3$ [44], and ScMnO$_3$ [45]. As argued in Ref. 17, the spin-reorientation transition at $T_{SR}$ in *h-R*MnO$_3$ family correlates strongly with the Mn position, i.e., the spin reorientation happens when x_Mn crosses the 1/3 threshold. But in both LuMn$_{0.7}$Fe$_{0.3}$O$_3$ and LuMn$_{0.5}$Fe$_{0.5}$O$_3$, no such cross has been observed at around $T_{SR}$ within the resolution of our NPD experiments (see Fig. 6(b)). In contrast to the NPD data of LuFe$_{1-x}$Mn$_x$O$_3$ ($x$ = 0.25 and 0.33) [23], those of LuMn$_{0.7}$Fe$_{0.3}$O$_3$ show no noteworthy change across $T_{SR}$ = 55 K as shown in Fig. 8(a). It is suggested in Fig. 7(e) that this spin reorientation is a gradual spin rotation taking place between the base temperature and $T_N$. Although the spin reorientation appears as a peak in magnetization only for LuMn$_{0.7}$Fe$_{0.3}$O$_3$ as seen in Fig. 2(a), this thermally-driven spin reorientation should also exist in LuMn$_{0.8}$Fe$_{0.2}$O$_3$ and LuMn$_{0.9}$Fe$_{0.1}$O$_3$. There is seemingly a very broad hump in the temperature dependence of magnetization of LuMn$_{0.8}$Fe$_{0.2}$O$_3$ and LuMn$_{0.9}$Fe$_{0.1}$O$_3$ at around 55 K, as compared with the curve of LuMnO$_3$. Therefore the spin reorientation at $T_{SR}$ can also be taken as an evidence of the drastic effect due to Fe-doping. A precise determination of the nature of the spin reorientation in these compounds requires detailed neutron investigations on single crystals.

A relatively complete understanding to the LuMn$_{1-x}$Fe$_x$O$_3$ series is achieved by combining the results of this work and Ref. 23, as summarized in the magnetic phase diagram in Fig. 9 as functions of temperature and the Fe concentration. The $T_N$ of $h$-LuFeO$_3$ and LuMn$_{0.5}$Fe$_{0.5}$O$_3$ are taken from Ref. 46 and Ref. 2, respectively. The magnetic phase of LuMn$_{1-x}$Fe$_x$O$_3$ for $0.3 < x < 0.5$ and $0 < x < 0.1$ below $T_N$ is still unclear, but it is reasonable to presume mixed spin presentations emerge as well in the magnetic ground states of these compositions. It is obvious that $T_N$ increases almost linearly with increasing Fe concentration. The magnetic transition temperature of $h$-LuMnO$_3$ is thus tunable by tailoring the transition-metal composition.

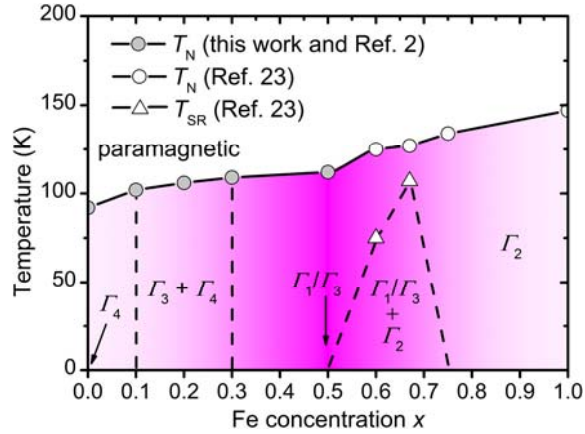

FIG. 9. (Color online) Magnetic phase diagram determined from neutron scattering measurements. The diagram for $x > 0.5$ is taken from Ref. 23. The $T_N$ and spin representation for $x = 0.5$ are taken from Ref. 2.

## V. CONCLUSION

We have studied the nuclear and magnetic structures of LuMn$_{1-x}$Fe$_x$O$_3$ ($x = 0, 0.1, 0.2$, and $0.3$) using magnetic measurements and neutron powder diffraction. The nuclear structures of the samples preserve the $P6_3cm$ space group of $h$-$R$MnO$_3$. The atomic positions undergo clear displacements at $T_N$, suggesting strong spin-lattice coupling. The magnetic ground state has been evaluated with a mixed spin configuration of $\Gamma_3$ and $\Gamma_4$ representations for the Fe-doped samples, and a single representation of $\Gamma_4$ for LuMnO$_3$. The short-range order has been evidenced by the diffuse neutron scattering and attributed to the strong exchange couplings in *a-b* planes. A composition-driven spin reorientation introduced by Fe-doping has been highlighted. A thermally-driven spin rotation from $\Gamma_3$ to $\Gamma_4$ has also been revealed in the Fe-doped samples as the temperature changes from the base temperature to $T_N$. It is confirmed that the AFM transition temperature of *h*-LuMnO$_3$ can be raised by Fe-doping.

## ACKNOWLEDGMENTS

We thank Dr. Kirill Nemkovski for his help in neutron scattering experiments. The work was partially performed at the neutron spallation source SINQ (PSI, Switzerland).

## REFERENCES

[1] R. D. Johnson and P. G. Radaelli, Annu. Rev. Mater. Res. **44**, 269 (2014) and references therein.

[2] H. S. Nair, Z. Fu, C. M. N. Kumar, V. Y. Pomjakushin, Y. Xiao, T. Chatterji, and A. M. Strydom, Europhysics Letters **110,** 37007 (2015).

[3] T. Katsufuji, M. Masaki, A. Machida, M. Moritomo, K. Kato, E. Nishibori, M. Takata, M. Sakata, K. Ohoyama, K. Kitazawa, and H. Takagi, Phys. Rev. B **66**, 134434 (2002).

[4] F. Yen, C. R. dela Cruz, B. Lorenz, Y. Y. Sun, Y. Q. Wang, M. M. Gospodinov, and C. W. Chu, Phys. Rev. B **71**, 180407(R) (2005).

[5] S. Lee, A. Pirogov, J. H. Han, J.-G. Park, A. Hoshikawa, and T. Kamiyama, Phys. Rev. B **71**, 180413(R) (2005).

[6] H. L. Yakel, W. C. Koehler, E. F. Bertaut, and E. F. Forrat, Acta Crystallogr. **16**, 957 (1963).

[7] T. J. Sato, S.-H. Lee, T. Katsufuji, M. Masaki, S. Park, J. R. D. Copley, and H. Takagi, Phys. Rev. B **68**, 014432 (2003).

[8] A. Muñoz, J. A. Alonso, M. J. Martínez-Lope, M. T. Casáis, J. L. Martínez, and M. T. Fernández-Díaz, Phys. Rev. B **62**, 9498 (2000).

[9] E. F. Bertaut and M. Mercier, Phys. Lett. A **5**, 27 (1964).

[10] J. Park, U. Kong, A. Pirogov, S.I. Choi1, J.-G. Park, Y. N. Choi, C. Lee, and W. Jo, Appl. Phys. A **74**, S796 (2002).

[11] M. Fiebig, D. Fröhlich, K. Kohn, St. Leute, Th. Lottermoser, V. V. Pavlov, and R. V. Pisarev, Phys. Rev. Lett. **84**, 5620 (2000).

[12] D. Fröhlich, St. Leute, V. V. Pavlov, and R. V. Pisarev, Phys. Rev. Lett. **81**, 3239 (1998).

[13] M. Chandra Sekhar, S. Lee, G. Choi, C. Lee, and J.-G. Park, Phys. Rev. B **72**, 014402 (2005).

14  S. Namdeo, S. S. Rao, S. D. Kaushik, V. Siruguri, and A. M. Awasthi, J. Appl. Phys. **116**, 024105 (2014).

15  P. J. Brown and T. Chatterji, J. Phys.: Condens. Matter **18**, 10085 (2006).

16  S. Lee, A. Pirogov, M. Kang, K.-H. Jang, M. Yonemura, T. Kamiyama, S.-W. Cheong, F. Gozzo, N. Shin, H. Kimura, Y. Noda, and J.-G. Park, Nature **451**, 805 (2008).

17  X. Fabrèges, S. Petit, I. Mirebeau, S. Pailhès, L. Pinsard, A. Forget, M. T. Fernandez-Diaz, and F. Porcher, Phys. Rev. Lett. **103**, 067204 (2009).

18  J. Park, S. Lee, M. Kang, K.-H. Jang, C. Lee, S. V. Streltsov, V. V. Mazurenko, M. V. Valentyuk, J. E. Medvedeva, T. Kamiyama, and J.-G. Park, Phys. Rev. B **82**, 054428 (2010).

19  I. V. Solovyev, M. V. Valentyuk, and V. V. Mazurenko, Phys. Rev. B **86**, 054407 (2012).

20  D. P. Kozlenko, S. E. Kichanov, S. Lee, J.-G. Park, V. P. Glazkov, and B. N. Savenko, JETP lett. **83**, 346 (2006).

21  S.L. Samal, W. Green, S.E. Lofland, K.V. Ramanujachary, D. Das, and A.K. Ganguli, Journal of Solid State Chemistry **181**, 61 (2008).

22  N. Sharma, A. Das, C. L. Prajapat, and S. S. Meena, J. Magn. Magn. Mater. **348**, 120 (2013).

23  S. M. Disseler, X. Luo, B. Gao, Y. S. Oh, R. Hu, Y. Wang, D. Quintana, A. Zhang, Q. Huang, J. Lau, R. Paul, J. W. Lynn, S.-W. Cheong, and W. Ratcliff II, Phys. Rev. B **92**, 054435 (2015).

24  T. Katsufuji, S. Mori, M. Masaki, Y. Moritomo, N. Yamamoto, and H. Takagi, Phys. Rev. B **64**, 104419 (2001).

25  P. Fischer, G. Frey, M. Koch, M. Könnecke, V. Pomjakushin, J. Schefer, R. Thut, N. Schlumpf, R. Bürge, U. Greuter, S. Bondt, and E. Berruyer, Physica B **276**, 146 (2000).

26  M. Hoelzel, A. Senyshyn, and O. Dolotko, Journal of large-scale research facilities **1**, A5 (2015).

27  A. S. Wills, Physica B **276**, 680 (2000).

28  H. M. Rietveld, J. Appl. Cryst. **2**, 65 (1969).

29  J. Rodriguez-Carvajal, Physica B **192**,55 (1993).

30  J. Park, J.-G. Park, G. S. Jeon, H.-Y. Choi, C. Lee, W. Jo, R. Bewley, K. A. McEwen, and T. G. Perring, Phys. Rev. B **68**, 104426 (2003).

31  J. Park, M. Kang, J. Kim, S. Lee, K.-H. Jang, A. Pirogov, and J.-G. Park, Phys. Rev. B **79**, 064417 (2009).

32  P. Tong, D. Louca, N. Lee, and S.-W. Cheong, Phys. Rev. B **86**, 094419 (2012).

33  B. B. Van Aken, A. Meetsma and T. T. M. Palstra, Acta Cryst. E**57**, i101 (2001).

34  H. Das, A. L. Wysocki, Y. Geng, W. Wu, and C. J. Fennie, Nature Communications **5**, 2998 (2014).

35  L. L. Lumata, T. Besara, P. L. Kuhns, A. P. Reyes, H. D. Zhou, C. R. Wiebe, L. Balicas, Y. J. Jo, J. S. Brooks, Y. Takano, M. J. Case, Y. Qiu, J. R. D. Copley, J. S. Gardner, K. Y. Choi, N. S. Dalal, and M. J. R. Hoch, Phys. Rev. B **81**, 224416 (2010).

36  A. S. Wills, N. P. Raju, C. Morin, and J. E. Greedan, Chem. Mater. **11**, 1936 (1999).

37  B. E. Warren, Phys. Rev. **59**, 693 (1941).

38  Z. Fu, Y. Zheng, Y. Xiao, S. Bedanta, A. Senyshyn, G. G. Simeoni, Y. Su, U. Rücker, P. Kögerler, and T. Brückel, Phys. Rev. B **87**, 214406 (2013).

39  Z.-D. Fu, P. Kögerler, U. Rücker, Y. Su, R. Mittal, and T. Brückel, New J. Phys. **12**, 083044 (2010).

40  H. S. Nair, Z. Fu, J. Voigt, Y. Su, and T. Brückel, Phys. Rev. B **89**, 174431 (2014).

41  Th. Brückel, C. Paulsen, W. Prandl, and L. Weiss, J. Phys. I France **3**, 1839 (1993).

42  D. P. Kozlenko, S. E. Kichanov, S. Lee, J.-G. Park, V. P. Glazkov, and B. N. Savenko, JETP lett. **82**, 193 (2005).

43  H. Wang, I. V. Solovyev, W. Wang, X. Wang, P. J. Ryan, D. J. Keavney, J.-W. Kim, T. Z. Ward, L. Zhu, J. Shen, X. M Cheng, L. He, X. Xu, and X. Wu, Phys. Rev. B **90**, 014436 (2014).

44  B. Lorenz, A. P. Litvinchuk, M. M. Gospodinov, and C. W. Chu, Phys. Rev. Lett. **92**, 087204 (2004).

45  M. Bieringer and J. E. Greedan, J. Solid State Chem. **143**, 132 (1999).

46  J. A. Moyer, R. Misra, J. A. Mundy, C. M. Brooks, J. T. Heron, D. A. Muller, D. G. Schlom, and P. Schiffer, APL Mater. **2**, 012106 (2014).